\documentclass[prb,preprint,showpacs,amsmath,amssymb,superscriptaddress]{revtex4}
\usepackage{epsfig}
\usepackage{graphicx}
\usepackage{amsmath,amssymb}

\newcommand{\beq}[1]{
\begin{equation}
\label{e#1} }

\newcommand{\eeq}{
\end{equation}
}
\newcommand{\rr}{{\bf r}}
\newcommand{\kk}{{\bf k}}
\begin{document}

\title{Spin-injection Hall effect in a planar photovoltaic cell}

\author{J.~Wunderlich}
\affiliation{Hitachi Cambridge Laboratory, Cambridge CB3 0HE, United Kingdom}
\affiliation{Institute of Physics ASCR, v.v.i., Cukrovarnick\'a 10, 162 53
Praha 6, Czech Republic}

\author{A.~C.~Irvine}
\affiliation{Microelectronics Research Center, Cavendish Laboratory,
University of Cambridge, CB3 0HE, United Kingdom}

\author{Jairo~Sinova}
\affiliation{Department of Physics, Texas A\&M University, College
Station, TX 77843-4242, USA}
\affiliation{Institute of Physics ASCR, v.v.i., Cukrovarnick\'a 10, 162 53 Praha 6, Czech Republic}

\author{B.~G.~Park}
\affiliation{Hitachi Cambridge Laboratory, Cambridge CB3 0HE, United Kingdom}

\author{X.~L.~Xu}
\affiliation{Hitachi Cambridge Laboratory, Cambridge CB3 0HE, United Kingdom}

\author{B.~Kaestner}
\affiliation{Physikalisch-Technische Bundesanstalt, Bundesallee 100, 38116 Braunschweig, Germany}

\author{V. Nov\'ak}
\affiliation{Institute of Physics ASCR, v.v.i., Cukrovarnick\'a 10, 162 53
Praha 6, Czech Republic}

\author{T.~Jungwirth}
\affiliation{Institute of Physics ASCR, v.v.i., Cukrovarnick\'a 10, 162 53
Praha 6, Czech Republic} \affiliation{School of Physics and
Astronomy, University of Nottingham, Nottingham NG7 2RD, United Kingdom}
\date{\today}
%
%\begin{abstract}
%We discuss the realization of a p-n junction transistor which allows for large hole depletions in (Ga,Mn)As thin ferromagnetic films at a few Volts.  The utility of this all-semiconductor ferromagnetic FET  in spintronic research is demonstrated by measuring field-effects on the anisotropic magnetoresistance, Curie temperature, and coercive fields.
%\end{abstract}
%

\maketitle

{\bf
Successful incorporation of the spin degree of freedom in semiconductor technology requires the development of a new paradigm allowing for a scalable, non-destructive electrical detection of the spin-polarization of injected charge carriers as they propagate along the semiconducting channel. In this paper we report the observation of a spin-injection Hall effect (SIHE) which exploits the quantum-relativistic nature of spin-charge transport and which meets all these key requirements on the spin detection. The two-dimensional electron-hole gas  photo-voltaic cell we designed to observe the SIHE allows us to develop a quantitative microscopic theory of the phenomenon and to demonstrate  its direct application in optoelectronics. We report an experimental realization of a non-magnetic spin-photovoltaic\cite{Zutic:2001_a} effect via the SIHE, rendering our device an electrical polarimeter which directly converts the degree of circular polarization of light to a voltage signal.\cite{Ganichev:2001_a}
}

Magneto-optical imaging\cite{Kikkawa:1999_a,Crooker:2005_a,Weber:2006_a,Lou:2007_a} of electron spins is an established qualitative probe which, although non-destructive, does not currently offer a sub-micron resolution and the possibility to integrate the probe within a semiconductor device. Methods introduced to date that allow for such an integration are destructive. One is based on extracting the spin-polarized current by a ferromagnet\cite{Hammar:2002_a,Lou:2007_a} and this implies further complications stemming from the semiconductor-magnet hybrid design. The other method is based on radiative electron-hole recombination\cite{Fiederling:1999_a,Ohno:1999_b,Zhu:2001_a,Jiang:2005_b} and here an additional device is needed to convert the emitted circularly polarized light into an electrical signal. The SIHE we introduce in our work is a transverse voltage response to the local spin-polarization of injected charge carriers. Its physical origin is related to the anomalous Hall effect (AHE) in uniformly polarized systems,\cite{Chazalviel:1975_a,Ohno:1992_a,Cumings:2006_a,Miah:2007_a}  i.e. is due to the spin-orbit coupling effects on electrical transport, but unlike the AHE is observed outside the area in which spin-polarization is generated and is spatially non-uniform. The SIHE is also distinct from the (inverse) spin Hall effects\cite{Kato:2004_d,Wunderlich:2004_a,Valenzuela:2006_a} which are based on the subtle concept of pure spin-currents.\cite{Sinova:2005_b} We demonstrate in this paper that the SIHE is a non-destructive local probe embedded directly along the semiconducting channel, operates without external magnetic fields or magnetic elements introduced in the structure, and has the resolution given by the state-of-the-art semiconductor processing limits. These experimental results combined with a detailed microscopic theory picture of the SIHE we provide imply broad implications of the effect for basic physics research and applications in the area of spin-charge dynamics phenomena.

The design of the all-semiconductor SIHE microchips utilized in our study combines conventional approaches to semiconductor device fabrication with processing steps dictated by the special requirements on the electrical spin detection. Molecular-beam epitaxy grown  (Al,Ga)As/GaAs heterostructure is the starting semiconductor material, chosen for its well established optical generation of spin-polarized electron-hole pairs. The  requirements to spatially separate electrons and holes  and to inject the respective spin-polarized currents into thin Hall bar channels  led us to utilize a  co-planar two-dimensional electron-hole gas (2DEG-2DHG) p-n junction, illustrated schematically in Fig.~1(a). The wafer, grown along the [001]-axis, consists of a modulation p-doped AlGaAs/GaAs heterojunction on  top of the structure separated by 90~nm of intrinsic GaAs from an n-doped AlGaAs/GaAs heterojunction underneath (for further details see "wafer 2" in Ref.~\onlinecite{Kaestner:2004_thesis} and Supplemetary material). In the unetched part of the wafer the top heterojunction is populated by the 2DHG  while the 2DEG at the bottom heterojunction is depleted. The n-side of the co-planar p-n junction is formed by removing the p-doped surface layer from a part of the wafer which results in populating  the 2DEG. The corresponding 2D electron density  is $n_{\rm 2DEG}=$2.5$\times 10^{11}$~cm$^{-2}$ with mobility  $\mu=3\times 10^3$~cm$^2$/Vs.

Detailed numerical simulations\cite{Kaestner:2004_thesis} of the structure show that the lateral extent of the depletion layer at the p-n junction and of the corresponding built-in potential is approximately 100~nm. This sets the upper limit on the width of the optical generation region of counter-propagating electron and hole currents at zero bias. Measurements with variable position of the $\sim 1$~$\mu$m laser spot at reverse bias and sub-gap wavelengths used in the experiments confirmed negligible photo-generation inside the 2DEG channel, certainly beyond a distance of 1~$\mu$m from the p-n junction.  (Note that the Stark effect allows for sub-gap excitations from or to the confined 2D states near the p-n junction due to the strong build in electric field.\cite{Wunderlich:2004_a}) In all measurements presented below, the 2DEG Hall probes n1-n3, placed at distances of 1.5, 3.5, and 5.5~$\mu$m from the p-n junction along the 1~$\mu$m wide [1$\bar{1}$0]-oriented Hall bar, are  therefore outside the optical spin-generation area.

The spin-orbit coupling length in our 2DEG, discussed in detail in the theory section below, is of a micrometer scale. In order to provide local detection of the spin-polarization, the width of the Hall contacts has to be therefore substantially smaller than a micrometer. We have optimized the electron-beam lithography and reactive ion etching procedures for our co-planar p-n junction wafer to achieve contact widths of approximately 100~nm, as shown in the scanning electron micrograph in Fig.~1(b). To guarantee proper alignment throughout all fabrication steps, the probe n0 placed next to the p-n junction is wider (750~nm) and can overlap with the spin-generation area. Voltages detected at this probe are interpreted as a combination of the AHE and of an averaged SIHE signal. For reference measurements, additional Hall probes were placed symmetrically in the 2DHG whose typical spin-orbit and spin-coherence lengths are below the limits of the state-of-the-art lithography. We have studied five devices fabricated from the wafer all showing reproducible SIHE characteristics.

Typical SIHE signals in the 2DEG (with the laser spot focused on the p-n junction) are demonstrated in Fig.~1(c) for  probe n2. Using a linear polarizer and a $\lambda/2$ wave-plate we prepare in step~1 a linearly polarized light beam at +45$^{\circ}$ angle with respect to the optical axis of a photo-elastic modulator (PEM). The PEM retardation is set to oscillate between $\pm 90^{\circ}$, producing a circularly polarized light with oscillating helicity. Hall voltages are measured by recording the amplitude and phase of the signal from a lock-in amplifier with a reference frequency of the  retardation oscillation of the PEM. In step~2, the PEM retardation is set to $0^{\circ}$ and in the middle of this measurement we manually rotate the $\lambda/2$ wave-plate so that the incoming linearly polarized light beam is at -45$^{\circ}$ angle with respect to the optical axis of the PEM. In step~3, the PEM retardation is reset to its state in step~1, i.e., the left and right circular polarizations of the beam incident on the sample switched places compared to step 1 (for further details on the experimemtal technique see Supplementary material). While the longitudinal resistance is insensitive to the polarization of the incident light, the Hall signals are antisymmetric with respect to the helicity of the  circular polarization. Due to the optical selection rules, the spin-polarization of the injected carriers is determined by the polarization of the incident light, i.e., the sign of the Hall signal we observe reverses upon reversing the out-of-plane spin-polarization of the optically generated conduction electrons. In the theory section we show that depending on the local out-of-plane component of their spin the electrons are asymmetrically deflected  towards the edges of the 2DEG channel due to spin-orbit coupling, creating a finite Hall voltage. The observation of the Hall signals at zero bias at probe n2 and the similarity between data at zero and -10~V bias  confirm, as explained above, that  Hall signals in these experiments are detected away from the generation area, i.e., we measure the SIHE.

Detailed dependencies of the SIHE at probes n1 and n2 on the degree of circular polarization of the incident light, as modulated by the PEM, are presented in Fig.~2(a) and (b). Here we plot the Hall angles determined as the ratio between longitudinal sheet resistance, measured between successive  Hall contacts, and transverse resistances. The SIHE signals are linear in the degree of polarization which is analogous to the proportionality of the AHE  to the magnetization in uniformly polarized systems. We also point out the large magnitude of the SIHE. The Hall angles  we observe are comparable to anomalous Hall angles in metal ferromagnets.

In Fig.~2(c) and (d) we present Hall measurements in the 2DHG. We compare  signals acquired with the laser spot focused on the p-n junction as in the case of the SIHE measurements in the 2DEG, and then moved towards the first 2DHG Hall probe next to the p-n junction and defocused. Consistent with the small spin coherence length in the strongly spin-orbit coupled 2DHG, no clearly measurable SIHE  is found with the laser spot focused on the p-n junction.  We observe, however, a strong transverse-voltage signal when uniformly  polarizing a larger area of the 2DHG around the Hall probe. This measurement links our work to previous studies of the AHE in homogeneously polarized semiconductors\cite{Chazalviel:1975_a,Ohno:1992_a,Cumings:2006_a,Miah:2007_a} and emphasizes the distinction between the AHE and SIHE. Unlike the AHE, the SIHE probes the spin of carriers outside the region subject to fields which generate spin-polarization.

Fig.~3(a) shows the SIHE measured at the 2DEG probe n2 at 100, 160, and 220~K (at temperatures where the p-n junction leakage current is still negligible), demonstrating that the effect is readily detectable at high temperatures. Together with the zero-bias operation demonstrated in Fig.~1(c) and linearity in the degree of circular polarization of the incident light shown in Fig.~2, these characteristics represent the realization of the spin-photovoltaic effect in a non-magnetic structure and demonstrate the utility of the device as an electrical polarimeter. Note that our approach is distinct from the former proposal of the spin-voltaic effect which assumes different longitudinal transport coefficients for spin-up and spin-down channels due to a non-zero equilibrium spin-polarization in the system.\cite{Zutic:2001_a,Kondo:2006_a}

In Fig.~3(b) we plot simultaneously the Hall signals measured in all available 2DEG probes. As an experimental consistency check, we skipped the measurement step~2 when acquiring these data. The manual rotation of the $\lambda/2$ wave-plate, causing a continuous rotation of the polarization angle of the incoming beam from +45$^{\circ}$ to -45$^{\circ}$ with respect to the  optical axis of the PEM, was performed with fixed PEM retardation setting.  The observed Hall signals again change sign for reversed helicities of the incident light. The data in Fig.~3(b) underline the capability of the SIHE to detect locally spin-polarization of injected electrical currents, which in general can be non-uniform along the conducting channel. Signals of one sign, for a given helicity of the incident light, are observed at probes n0 and n2 and signals of opposite sign are detected at probes n1 and n3. Recall that the  probe n0 covers a relatively large area which together with the non-uniformity of the  spin polarization explains smaller voltage  measured at this probe compared to probe n2, despite the immediate vicinity of probe n0 to the p-n junction. The alternation of the sign of simultaneously detected  SIHE signals (see also Figs.~2(a) and (b)) is another experimental observation which allows us to exclude spurious origin of the measured voltages. It also points towards the  utility of the SIHE in the research of coupled charge and spin dynamics of carriers injected into non-magnetic systems\cite{Schliemann:2003_c,Bernevig:2006_a,Weber:2006_a} and of the related spintronic device concepts.\cite{Datta:1990_a,Ohno:2008_a} These studies are, however, beyond the scope of our initial report of the SIHE. In the remaining paragraphs we keep the focus on the effect itself and establish our interpretation of the measured Hall signals by developing a semi-quantitative microscopic theory of the SIHE assuming parameters of the experimental system.

Our theoretical approach is based on the observation that the micrometer length scale governing the spatial dependence of the non-equilibrium spin-polarization  is much larger than the $\sim 20$~nm mean-free-path in our 2DEG  which governs the transport coefficients. This allows us to first calculate the steady-state spin-polarization profile along the channel and then consider the SIHE as a response to the local out-of-plane component of the polarization, as illustrated schamatically in Fig.~4(a) (see also Supplementary material for detailed theory derivations). The calculations start from the electronic structure of GaAs whose conduction band near the $\Gamma$-point is formed dominantly by Ga $s$-orbitals. This implies weak spin-orbit coupling originating from the mixing of the valence-band $p$-orbitals and from the broken inversion symmetry in the zincblende lattice. In the presence of an electric potential the corresponding 3D spin-orbit coupling Hamiltonian reads
\begin{equation}
H_{3D-SO}=\left[\lambda^* \boldsymbol\sigma \cdot(\kk\times{\bf \nabla}V(\rr))\right]+\left[{\cal B} k_x(k_y^2-k_z^2)\sigma_x+{\rm  cyclic \,\,\,permutations}\right]\,,
\label{3d-so}
\end{equation}
where $\boldsymbol\sigma$ are the Pauli spin matrices, $\kk$ is the momentum of the electron, ${\cal B}\approx 10$~eV\AA$^3$\;, and $\lambda^*=5.3$\AA$^2$\; for GaAs.\cite{Knap:1996_a,Winkler:2003_a}
Eq.~(\ref{3d-so}) together with the 2DEG confinement yield an effective 2D Rashba and Dresselhaus spin-orbit coupled Hamiltonian,\cite{Schliemann:2003_c,Bernevig:2006_a}
\begin{equation}
H_{\rm 2DEG}=\frac{\hbar^2 k^2}{2m}+\alpha(k_y\sigma_x-k_x\sigma_y)+\beta(k_x\sigma_x-k_y\sigma_y),
\end{equation}
where  $m=0.067 m_e$, $\beta=-{\cal B} \langle k_z^2 \rangle\approx -0.02$~eV\AA\; and $\alpha=e\lambda^* E_z\approx0.01-0.03$~eV\AA\; for the  strength of the confining electric field, $eE_z\approx 2-5\times 10^{-3}$ eV/\AA\;, obtained from a self-consistent Poisson-Schr\"{o}dinger simulation of the conduction band profile of our GaAs/AlGaAs heterostructure.\cite{Kaestner:2004_thesis,Wunderlich:2004_a}

In the weak spin-orbit coupling regime of our structure with $\alpha k_F$ and $\beta k_F$ ($\sim$0.5 meV) much smaller than the disorder scattering rate $\hbar/\tau$ ($\sim$5 meV), the system obeys a set of spin-charge diffusion equations.\cite{Bernevig:2006_a} In the steady-state we obtain that the spatial dependence of the out-of-plane component of the spin polarization along the [1$\bar{1}$0] channel direction is given by a damped oscillatory function $p_z(x_{[1\bar{1}0]})=\exp(qx_{[1\bar{1}0]})$ with the complex  wavevector $q=|q|\exp(i\theta)$, where
%\begin{equation}
$|q|=\left(\tilde{L}^2_1\tilde{L}^2_2+\tilde{L}^4_2\right)^{1/4}$,
$\theta =\frac{1}{2}\arctan\left( \frac{\sqrt{2\tilde{L}^2_1\tilde{L}^2_2-\tilde{L}_1^4/4}}{\tilde{L}^2_2-\tilde{L}^2_1/2}\right)$,
%\end{equation}
and $\tilde{L}_{1/2}=2m|\alpha\pm\beta|/\hbar^2$.

From the known local spin-polarization we calculate the Hall signal by realizing that the dominant contribution in the weak spin-orbit coupling regime is the extrinsic skew-scattering. This contribution  is obtained by considering asymmetric scattering from a spin-orbit coupled impurity potential originating from the first term in Eq.~(\ref{3d-so}).\cite{Nozieres:1973_a,Crepieux:2001_b} Within the second-order Born approximation for short-range scatterers we obtain the spatially dependent SIHE angle,\cite{Nozieres:1973_a,Crepieux:2001_b}
\begin{equation}
\alpha_H(x_{[1\bar{1}0]})=2\pi\lambda^*\sqrt{\frac{e}{\hbar n_i\mu}}\,n\,p_z(x_{[1\bar{1}0]})\;,
\label{alpha_H}
\end{equation}
where $n$ is the density of optically injected carriers into the 2DEG channel. In Fig.~4(b) we plot the resulting theoretical $\alpha_H$ along the  [1$\bar{1}$0] direction for the relevant range of Rashba and Dresselhaus parameters corresponding to our experimental structure. We have assumed a donor impurity density $n_i$ of the order of the equilibrium density $n_{\rm 2DEG}$ of the 2DEG in dark, which is an upper bound for the strength of the impurity scattering in our modulation-doped heterostructure and, therefore, a lower bound for the Hall angle. For the mobility $\mu$ of the injected electrons in the 2DEG channel we considered the experimental value determined from ordinary Hall measurements without illumination. The density of photoexcited carriers of $n\approx 2\times 10^{11}$~cm$^{-2}$ was obtained from  the measured longitudinal resistance between successive Hall probes under illumination assuming constant mobility.

The theory results shown in Fig.~4(b) provide a semi-quantitative account of the magnitude of the observed SIHE angle ($\sim 10^{-3}$) and explain the linear dependence of the SIHE on the degree of spin-polarization of injected carriers. The calculations are also consistent with the experimentally inferred  precession length of the order of a micrometer and the spin-coherence exceeding micrometer length scales. We emphasize that the 2DEG in the strong disorder, weak spin-orbit coupling regime realized in our experimental structures is a particularly favorable system for theoretically establishing the presence of the SIHE. In this regime and for the simple band structure of the archetypal 2DEG, the spin-diffusion equations and the leading skew-scattering mechanism of the spin-orbit coupling induced Hall effect are well understood areas of the physics of quantum-relativistic spin-charge dynamics. The possibility to observe the SIHE in normal semiconductors, established in our work,  and to tune independently the strengths of disorder and spin-orbit coupling in semiconductor structures open new opportunities for resolving long-standing debates on the nature of spin-charge dynamics in the intriguing strong spin-orbit coupling regime. Apart from these basic physics problems, the SIHE can be directly implemented in devices such as the spin-photovoltaic cell demonstrated in this paper, in the Datta-Das\cite{Datta:1990_a} and related semiconductor spintronic transistors, and in a number of other microelectronic devices utilizing the spin degree of freedom of charge carriers.

We acknowledge support from EU Grant IST-015728, from Czech Republic Grants FON/06/E001, FON/06/E002, AV0Z10100521, KAN400100652, LC510, and Preamium Academiae, and from U.S. Grants  ONR-N000140610122,  DMR-0547875, and SWAN-NRI. Jairo Sinova is a Cottrell
Scholar of Research Corporation.

%\bibliographystyle{nature}
%\bibliography{MSWEBpublications,other}

\begin{figure}[h]
%\hspace*{-1cm}\epsfig{width=1.2\columnwidth,angle=0,file=fig1.pdf}
\hspace*{-1cm}\epsfig{width=0.8\columnwidth,angle=90,file=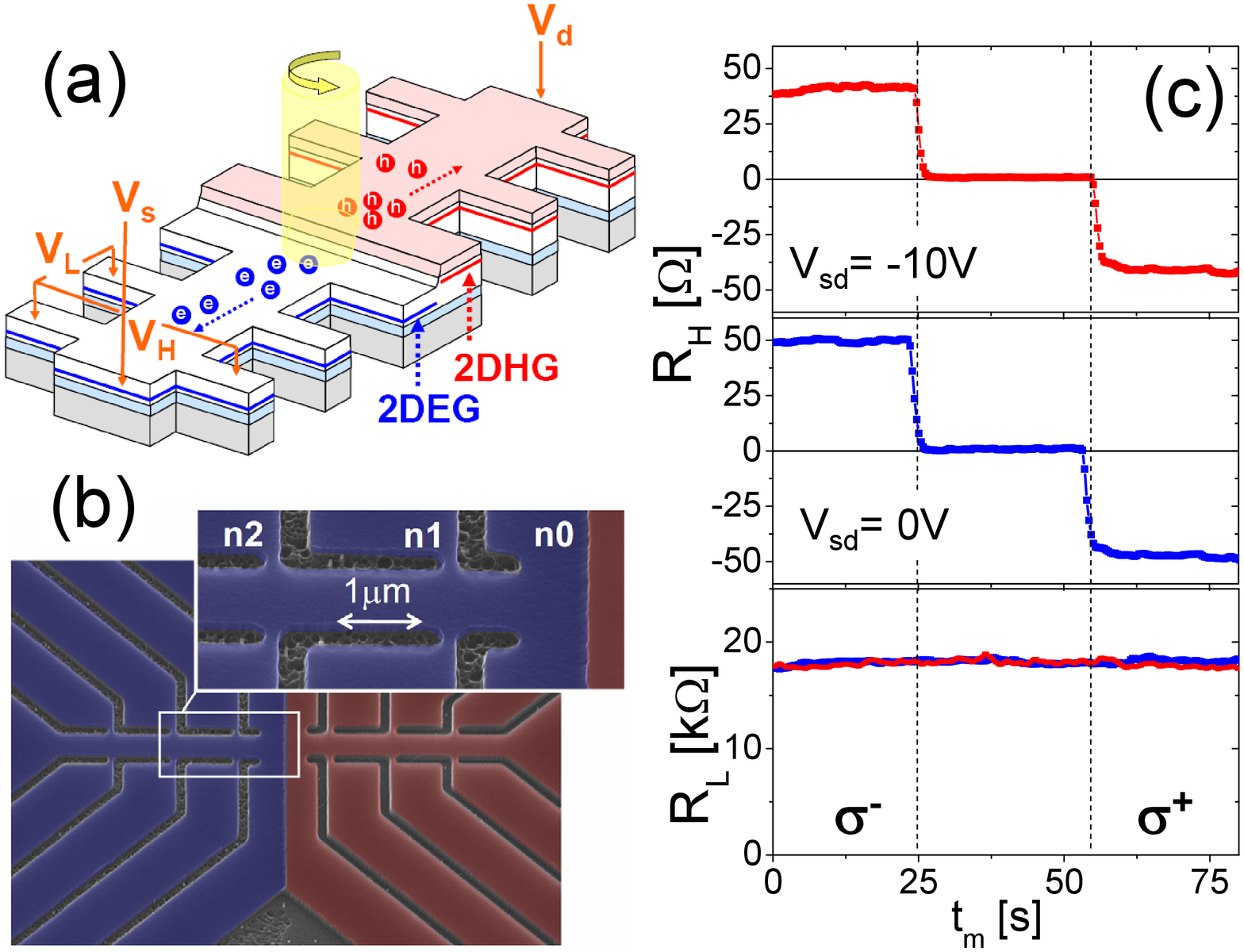}
\vspace*{0.5cm}
\caption{ (a) Schematic diagram of the co-planar 2DEG-2DHG p-n junction lithographically precessed to incorporate a set of Hall probes in the 2D electron and hole channels.(b) Scanning electron micrograph of the device. (c) Steady-state SIHE signals corresponding to opposite helicities of the incident light beam measured at the 2DEG Hall probe n2 at a laser wavelength of 850~nm, zero and reverse bias of -10~V, and 4~Kelvin. For the detail settings of the $\lambda/2$ wave-plate and the PEM during the measurements see the text.
}
\label{f1}
\end{figure}
%\pagebreak
\begin{figure}[h]
%\hspace*{-1cm}\epsfig{width=1.2\columnwidth,angle=0,file=fig1.pdf}
\hspace*{-2cm}\epsfig{width=0.9\columnwidth,angle=90,file=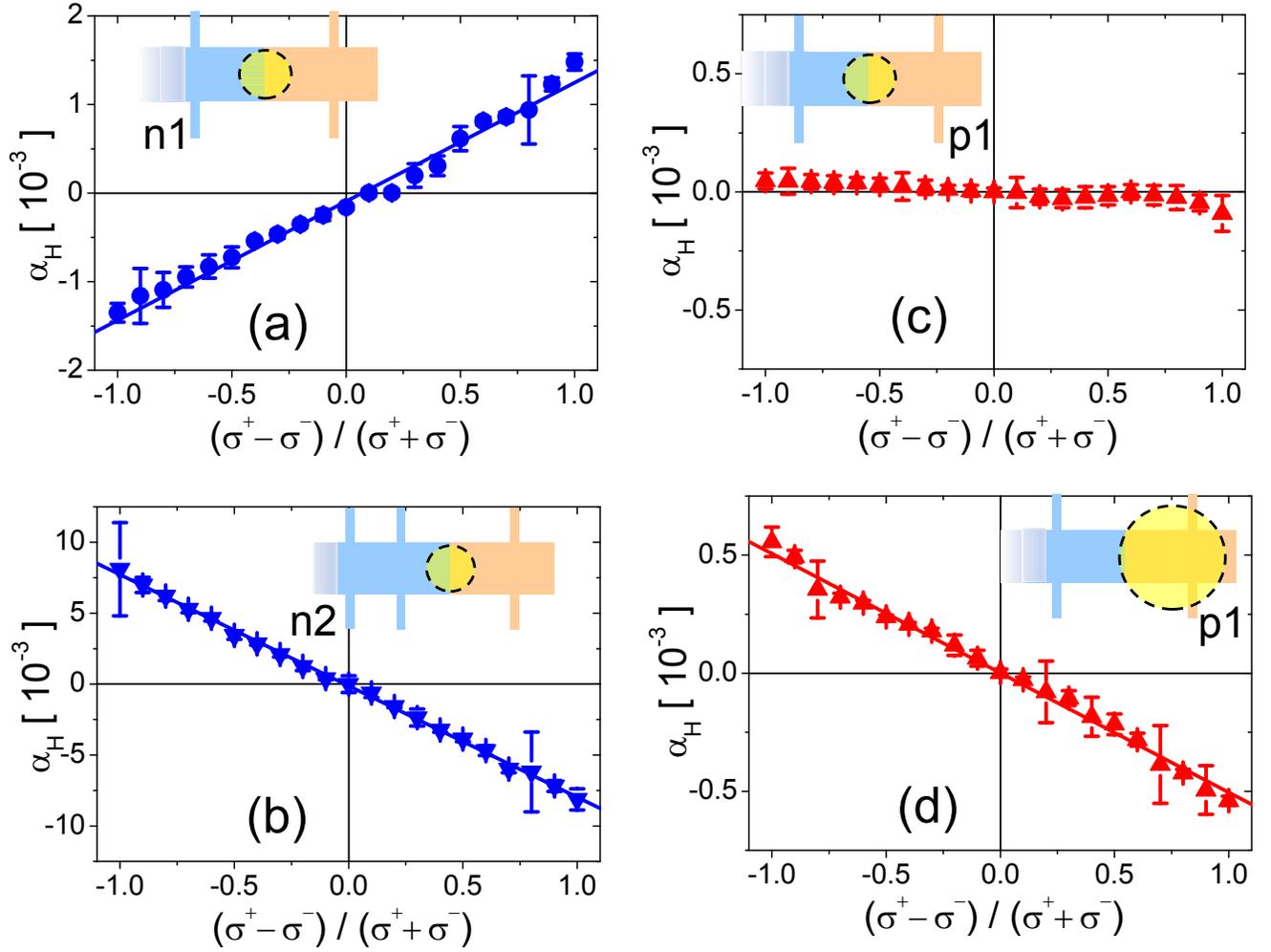}
\vspace*{0.5cm}
\caption{Experimental Hall angles, defined as the ratio between longitudinal sheet resistance measured between first and second  2DEG Hall contacts, and transverse resistances at the first (a) and second (b) 2DEG Hall probe. The SIHE angles are linear in the degree of  circular polarization of the incident light beam. (Each data point comprises integration over 2 minutes of measurement time.) The laser spot is focused on the p-n junction and also the other optical excitation conditions are the same as in Fig.~1. Measurements on the 2DHG Hall probe with the laser spot focused on the p-n junction is shown in panel (c) while panel (d) shows data for the spot moved towards the 2DHG probe and defocused. A measurable signal, corresponding to the AHE in a uniformly polarized 2DHG, is detected only in the latter case.
}
\label{f2}
\end{figure}

%\pagebreak
\begin{figure}[h]
%\hspace*{-1cm}\epsfig{width=1.2\columnwidth,angle=0,file=fig1.pdf}
\hspace*{-2cm}\epsfig{width=0.7\columnwidth,angle=90,file=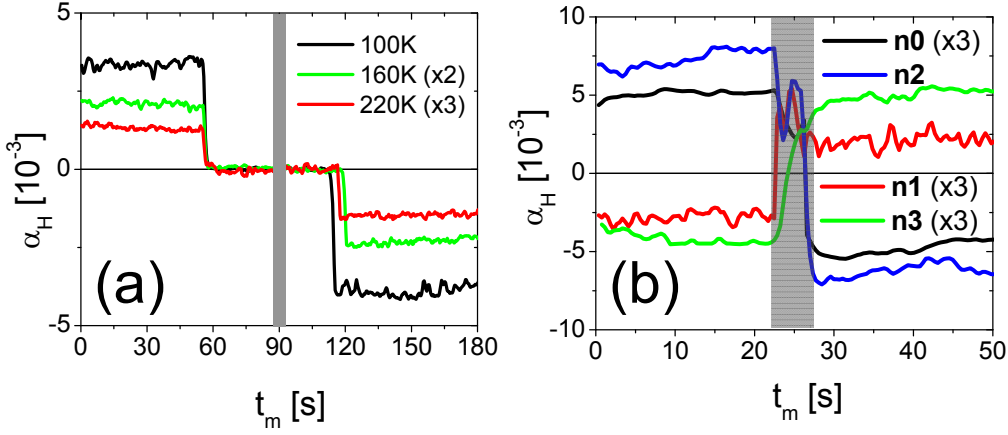}
\vspace*{-2cm}
\caption{(a) Steady-state SIHE measurements at the 2DEG probe n2 at 100, 160, and 220~K. These data together with the linear SIHE characteristics in Fig.~2(a) and (b) and zero bias operation shown in Fig.~1(c) demonstrate the realization of a transverse-voltage spin-photovoltaic effect without any magnetic elements in the structure or an applied magnetic field, and a functionality of the device as a solid-state electrical polarimeter. By choosing different semiconductor materials and heterostructures the device can operate over a wide range of infrared and visible light wavelengths. (b) SIHE signals recorder at a fixed PEM setting are plotted for all four Hall probes available in the 2DEG. Grey regions an (a) and (b) correspond to the manual rotation of the $\lambda/2$ wave-plate during the measurement.
}
\label{f3}
\end{figure}

%\pagebreak
\begin{figure}[h]
%\hspace*{-1cm}\epsfig{width=1.2\columnwidth,angle=0,file=fig1.pdf}
\hspace*{-2cm}\epsfig{width=0.9\columnwidth,angle=90,file=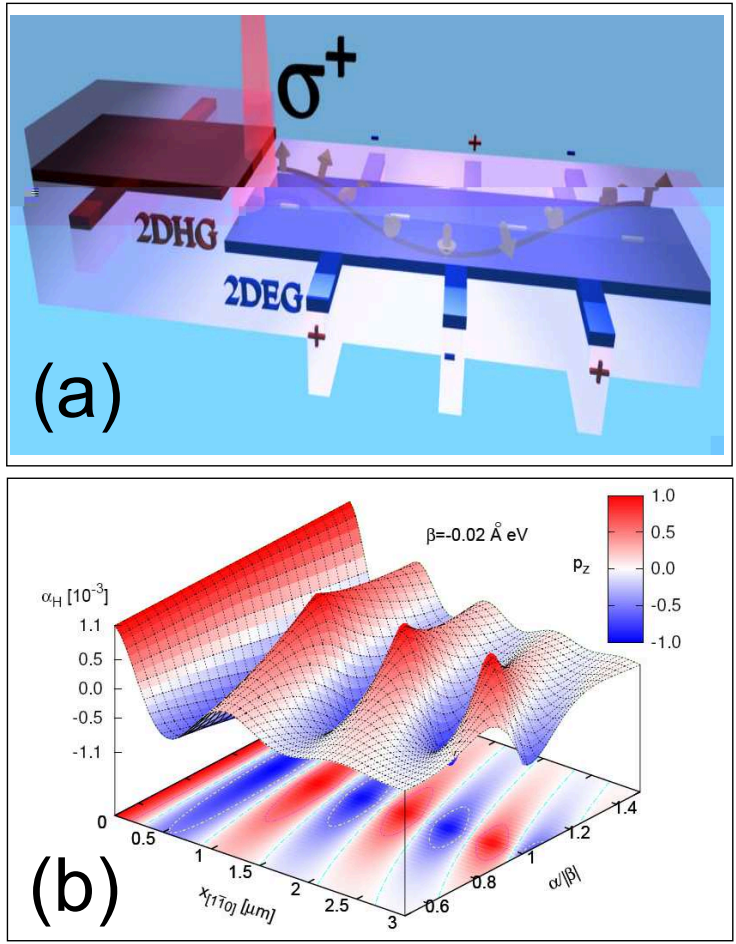}

\vspace*{0.5cm}
\caption{(a) Schematics of the co-planar spin-photovoltaic cell and of the origin of the SIHE. Spin-polarized electrons are injected from the p-n junction into the 2DEG channel. Electrical {\em in situ} detection  of the non-uniform spin polarization of moving electrons is enabled by the spin-dependent deflection of  electrons towards the channel edges, resulting in spatially varying Hall voltages. (b) Microscopic theory of the SIHE assuming spin-orbit coupled band structure parameters of the experimental 2DEG system. The calculated spin-precession and spin-coherence lengths and the magnitude of the Hall angles are consistent with experiment. The color-coded surface shows the proportionality between the Hall angle and the out-of-plane component of the spin-polarization.
}
\label{f4}
\end{figure}

\end{document}